\documentstyle[twoside,fleqn,psfig_sb]{an-art}
{A.~D.~Schwope and S. Mengel:  The eclipsing polar EP Draconis (=H1907+690)}
\begin{document}
\def\address #1END {{\vspace{9mm}\noindent\small Address of the authors: \medskip \\ #1}}
\def\addresses #1END {{\vspace{9mm}\noindent\small Addresses of the authors: \medskip \\ #1}}
\def\ep{EP Dra}
\def\pe{$\phi_{\rm ecl}$}
\def\kmps{km\,s$^{-1}$}
\def\msun{M$_{\sun} $}
\def\mwd{$M_{\rm wd}$}
\def\he2{He{\sc ii}\,$\lambda$4686}
\hyphenation{couple com-pa-ri-son re-pre-sent closing}
\begin{titlepage}
%%%%% The number of the first page of the paper
\setcounter{page}{1}
\def\makeheadline{\vbox to 0pt{\vskip -30pt\hbox to 50mm
{\small Astron. Nachr. 318 (1997) 1, 1--9 \hfill}}}
\makeheadline
%%%%% \title {Title of paper}
\title {Phase-resolved spectroscopy and photometry of the eclipsing polar EP 
Draconis (=H1907+690)}
%%%%% \author {{\sc Initials and Surname}, Cite, Country \\
\author{{\sc A.~D.~Schwope\footnote{Visiting Astronomer, German-Spanish
Astronomical Center, Calar Alto, operated by the Max-Planck-Institute f\"{u}r
Astronomie  jointly with the Spanish National Comission for Astronomy}
\ {\rm and} S.~Mengel},
Potsdam-Babelsberg, Germany \\
\medskip
%%%%% {\small Institute} \\
{\small Astrophysikalisches Institut Potsdam} \\
\bigskip
%%%%% Further authors:
% {\sc }, \\
% \medskip
% {\small }
}
\date{Received 1996 November 5; accepted } 
\maketitle
%
%%%%%\summary
\summary
We present phase-resolved optical spectroscopy and CCD photometry 
of the faint eclipsing polar EP Dra (H1907+690). A revised ephemeris is
derived which connects all 32000 binary cycles since its discovery 
by Remillard et al.~(1991). We found no difference between spin and
and orbital periods of the white dwarf. Changes in the light curve morphology
are attributed to a different beaming behaviour which might change on 
timescales as short as one or several orbital periods. Optical
light curve modelling was used to estimate the co-latitude of the 
accretion spot, which must be larger than $40^\circ$.
We have detected Zeeman absorption lines 
of H$\alpha$ originating in an accretion halo in a field of 16\,MG.
The low-resolution spectra reveal no indication
of resolved cyclotron harmonics, which is also suggestive of a relatively 
low field strength in the accretion region.
The Balmer emission lines contain  significant contributions from
the UV-illuminated hemisphere of the companion star, whereas the 
HeII\,$\lambda$4686   
emission originates predominantly from the accretion stream. 
The emission lines have a multi-component structure and we could single out
a narrow emission line in the H$\beta$ and H$\gamma$ lines. 
Its radial velocity amplitude suggests a low mass for the white dwarf,
if the lines are 
interpreted as being of reprocessed origin from the whole illuminated
hemisphere of the companion star.
END
%%%%% \keyw
%%%%% Text of keywordsEND
\keyw
Accretion -- cataclysmic variables -- AM Herculis binaries -- 
             stars: \ep\ -- stars: eclipsing -- stars: magnetic fieldsEND
%%%%% \AAAcla
%%%%% Number of subject classificationEND 
%%%%% (For classification see Astronomy and Astrophysics Abstracts)
\AAAcla
116; 117; 119END
\end{titlepage}

\kap{Introduction}
EP Draconis belongs to the rare species of eclipsing AM Herculis binaries
(`polars'), of which less than ten are known to date. 
Discovered as optical counterpart of the HEAO-1 source 
1H\,1903+689 by Remillard et al.~(1991, henceforth Rea91, who referred to it
as H1907+690 because of a mismatch of the true position 
with the HEAO-1 error box), it attracted 
only little attention, mainly due to its faintness at optical wavelengths,
$V \simeq 18^{\rm m}$, and its short orbital period, $P_{\rm orb} = 104.6$\,min. 
The combination of both inhibited to collect detailed observational data
with high phase resolution.
Only one follow-up observation with the PSPC (position sensitive proportional
counter) onboard ROSAT is reported by Schlegel and Mukai (1995).
These latter observations showed the characteristic two-component spectrum of
polars, consisting of a soft blackbody (10--15\,eV) and  a hard bremsstrahlung
component. The eclipse was not clearly resolved at X-ray wavelengths, again 
due to the faintness of the system, i.e.~low count rate, 
and a possible accumulated phase uncertainty at the 
time of the ROSAT observation. Hence, an accurate orbital period
is necessary in order to reach an unambigous interpretation of the 
X-ray data.

Eclipsing polars potentially offer good diagnostic opportunities for studies
of the accretion geometry (and changes therein) and  for studies of the stellar 
components of the system. We, therefore, have undertaken some efforts
to improve our understanding of this system in particular and  with it of the 
polars in general. Information about e.g.~spin-orbit coupling, masses of the 
stars, magnetic field strength of the white dwarf, origin of line emission,
which are of importance for the class as a whole, 
can be drawn only from studies of individuals and  will be addressed 
in this paper. 

The photometric observations of \ep \ served also as a test case for the 
capabilities of the renovated 70cm-telescope at the Astrophysical 
Institute Potsdam, 
which is located between the capitals of Berlin and  
Potsdam, hence, under a `beautifully' illuminated sky. 

We describe our observations in Sect.~2, present the analysis of the 
light curves and  the low-resolution spectra in Sects.~3 to 5 and  close the 
paper with a short discussion of the main results in Sect.~6.

\kap{Observations of \ep }
\hspace*{-7pt}\sect{Low-resolution optical spectroscopy}
\ep \ was observed with the 3.5m-telescope and  Cassegrain double-beam 
spectrograph (TWIN) at Calar Alto, Spain, during two nights in August 1992
for each one orbital cycle.
The exact dates of the observations are given in Table~1 %\ref{tabspec}, 
together with 
information about the spectral and  time resolution achieved.
Inside the TWIN the beam is split by a dichroic 
at $\sim$5500\,\AA \ and  fed into two separate
spectrographs. These were equipped with low-resolution gratings and 
large-format CCDs (RCA $15\mu m$ at the blue arm, GEC $22.5\mu m$ at 
the red arm,
RCA binned by 2 in spatial direction).
We thus were able to cover the whole optical range at 
fairly good resolution with one shot. One drawback of the TWIN
at that time was the relative long time needed for read-out and  data 
storage, which resulted in an overhead of 2.5\,min between successive 
exposures. 
This impacts strongly the phase-resolution of this 105\,min binary
or alternatively the signal-to-noise
ratio achievable in individual spectra. 
In order to maximize the observing efficiency 
we used two different approaches in the two nights. In the first night we
obtained {\it stepped} spectra where the object is jumped along the 
spectrograph slit while exposing (we collected spectral information 
at 10 positions in a 6300\,sec exposure), in the second night we obtained
{\it trailed} spectrograms with the object trailing along the slit while
exposing. We exposed for 7200\,sec with a trailing rate of 25\arcsec/h, which
resulted in phase resolutions as given in Tab.~1.%\ref{tabspec}. 
\bigskip

\begin{quote}
\aba
Table 1.
Spectroscopic observations of \ep 

\abe

\medskip

{\footnotesize
\begin{tabular}{lccccccc}
\hline\\[-1mm]
date    &  time & $D^{\rm -1}$ & spectral range $\Delta\lambda$ & spectral resol. & spectra & time/spect. & phase resol. \\
Y/M/D &  [UT] & \AA/mm       &       \AA       &  FWHM \AA      & number  & sec      & \\[2mm]  
\hline\\[-1mm]
92/08/19 & 21:10 -- 22:55 & 144 & 3330 -- 5510 & 6 &    10 & 630 & 0.1 \\
92/08/19 & 21:10 -- 22:55 & 160 & 5590 -- 9700 & 7--8 & 10 & 630 & 0.1 \\
92/08/21 & 19:57 -- 21:57 & 144 & 3330 -- 5510 & 6 &    46 & 156 & 0.025 \\
92/08/21 & 19:57 -- 21:57 & 160 & 5590 -- 9700 & 7--8 & 61 & 118 & 0.019 \\[2mm]
\hline
\end{tabular}
}
\end{quote}
\bigskip

The observations
were performed under good weather conditions with medium seeing of about 
2\arcsec. Observations of \ep \ were accompanied by taking bias-, dark-
and flatfield frames. Helium-Argon arc line spectra were obtained each
before and  after the trailed and  stepped spectra were taken. We observed
a number of spectrophotometric standard stars during the observation run,
which were used in order to flux-calibrate the spectra of \ep. They also 
provided background light sources suitable for determining the atmospheric
absorption (O$_2$, H$_2$O) in the near infrared. 

The long exposure time used prevented a complete subtraction of the night
sky lines, in particular in the red trailed spectrum. Neither linear nor 
high-order polynomial interpolation of the sky intensity above and  below
the region of the CCD exposed to \ep \ succeeded in a complete removal of the
intense sky lines. This turned out to be due to our imperfect knowledge 
of the flatfield and  reduced the usefulness of the red spectra longward
of $\sim$7500\,\AA. Another influence on the trailed spectra comes from 
slit losses due to imperfect guiding during the last third of the exposure 
caused by problems with the $\delta$-encoder of the telescope. This influences
mainly observations of \ep \ during the faint part of its orbital cycle
and thus has no major impact on the analysis.

With the given setup of the grating angles 
we were left with a short uncovered wavelength interval between the 
blue and  red spectra.

\sect{Differential photometry}
Differential photometry of \ep \ was obtained during several nights in May and  November 1995 with
the 70cm reflector of the Astrophysical Institute Potsdam. The telescope
was equipped with a large-format Tek-CCD with low read noise.
All exposures of \ep \ 
were taken through a broad-band Johnson V-filter. 
These measurements were influenced by some transparency
variations through clouds. The counts of \ep,  which were measured through 
software-generated apertures, were, therefore, reduced to those of nearby field
stars in the $8\times 8$\,arcmin$^2$ field-of-view. 
In one of the nights a field with photometric standard stars was
observed. We could thus derive a crude photometric calibration with 
an uncertainty of about 0.3 mag. Some more 
details of the observations are given
in Tab.~2.%\ref{tabphot}.
\bigskip

\begin{quote}
\aba
Table 2.
Photometric observations of \ep 

\abe

\medskip

{\footnotesize
\begin{tabular}{lccrc}
\hline\\[-1mm]
date    &  time & filter & exposures & integration time \\
Y/M/D   &  [UT] &        &           & sec \\[2mm]  
\hline\\[-1mm]
95/05/04 & 23:55 -- 0:43  & V &  9 & 240 \\
95/11/20 & 2:27 -- 4:24   & V & 35 & 180 \\
95/11/21 & 16:53 -- 21:10 & V & 107 & 120 \\
95/11/25 & 18:25 -- 21:57 & V & 67 & 150 \\
95/11/26 & 19:24 -- 23:18 & V & 52 & 150 
\\[2mm]
\hline
\end{tabular}
}
\end{quote}

\begin{figure*}[t]
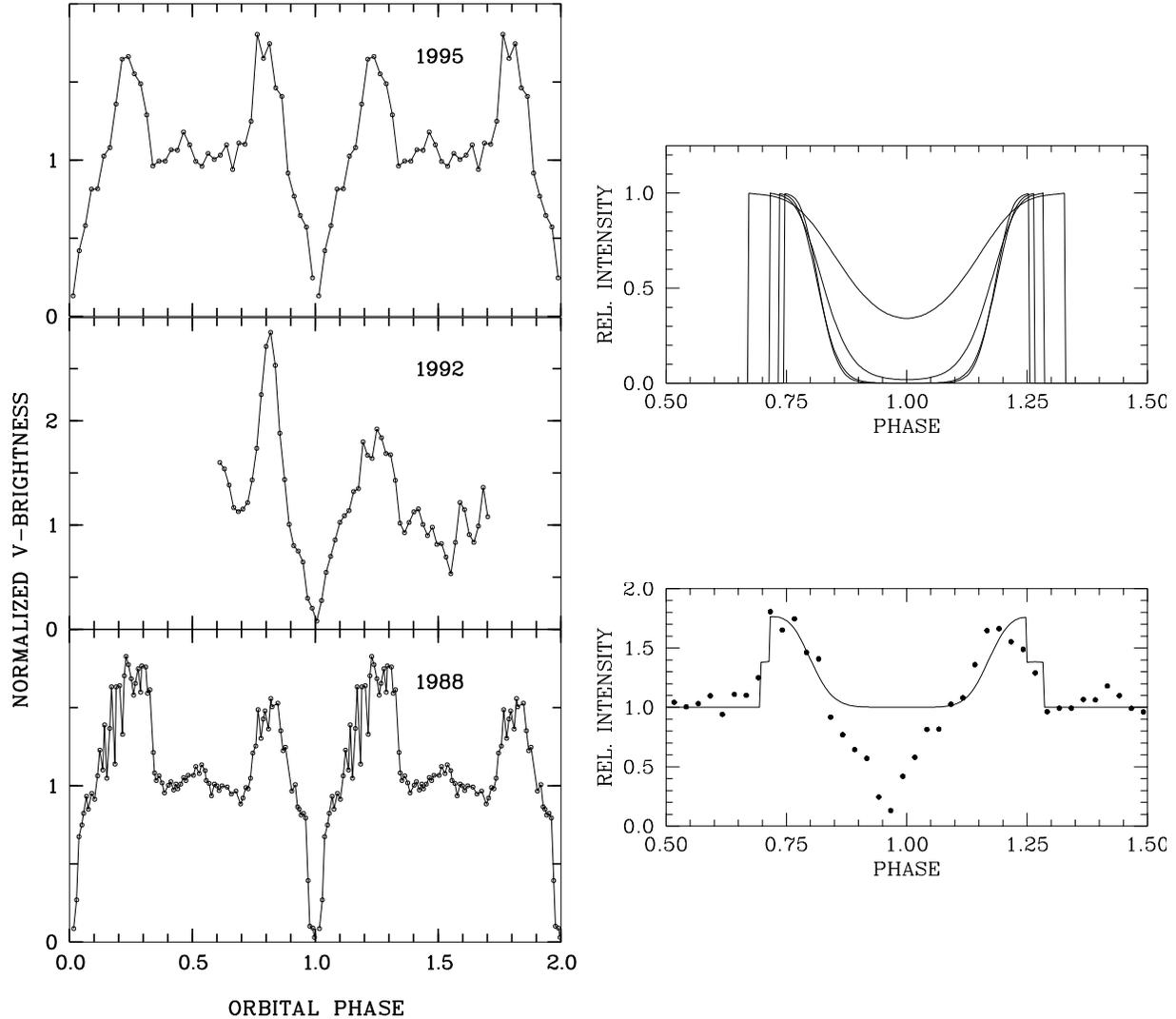

\begin{minipage}[]{188mm}
\begin{center}
\begin{minipage}[]{80mm}
\end{minipage}
\begin{minipage}[b]{15mm}
\end{minipage}
\begin{minipage}[]{80mm}
\vspace{2cm}
\end{minipage}
\end{center}
\end{minipage}

\caption[]{\label{ep_lcs} {\it (a, left)}
Optical V-band light curves of \ep\ obtained at different occasions 
during recent years ({\it top:} AIP 70cm, {\it middle:} trailed spectrogram,
{\it bottom:} data depicted from Rea91). 
All light curves were normalized to an approximate
brightness level of 1 at phase \pe = 0.5. The data of 1995 and  1988 were 
phase folded, the light curve derived from the trailed spectrogram is shown 
in original time sequence.
{\it (b, top right)} Synthesized light curves of a cyclotron emitting 
accretion spot with zero extent. The orbital inclination was fixed at
$i=80^\circ $, the accretion spot colatitude was varied in steps of $20^\circ $
between $20^\circ $ (longest visibility) and  $80^\circ $ (shortest visibility).
The phase convention defines minimum polar angle $\Theta$ as phase zero.
{\it (c, bottom right)}
Synthetic light curve assuming two cyclotron spots compared with the
1995 observations. 
}
\end{figure*}

\kap{Light curves and  eclipse ephemeris}
In Fig.~\ref{ep_lcs} we show the light curves of \ep \ as derived from 
the trailed spectrogram in the red arm of the TWIN and  from our
CCD photometry in comparison with the V-band light curve of Rea91. The
light curve derived from the trailed spectrogram was computed by averaging 
over the wavelength interval 5650--5740\AA \ and  thus roughly corresponds
to the standard V-band. At all occasions \ep\ displayed a similar light curve,
which is characterized by two intensity peaks at start and  end 
of the bright phase and  the eclipse due to transit of 
the secondary star inbetween. The average faint-phase brightness in 
1992 and  1995 was $V\simeq
18^{\rm m}$, maximum brightness in the peaks was $V \simeq 17^{\rm m}$.

We have measured the times of mid-eclipse in
all our photometric and  spectroscopic data (compiled in Tab.~3).
%\ref{ecl_tim}). 
An unweighted linear ephemeris to our new data combined with those 
measured by Rea91 gives an updated ephemeris for the time $T_0$ 
of mid-eclipse of

\begin{equation}
\mbox{ HJD} 
(T_0) = 244\,7681.72918(6) + E \times 0.072656259(5)\ ,
\end{equation}
where numbers in parenthesis indicate uncertainties in the last digits.
Phases given in this paper refer to this updated ephemeris. The 
inclusion of a quadratic term in the regression did not improve the fit.
\bigskip

\begin{quote}
\aba
Table 3.
Times of mid-eclipse of \ep\ from this work with their uncertainties

\abe

\medskip

{\footnotesize
\begin{tabular}{rcrc}
\hline\\[-1mm]
HJD & $\Delta T$ & cycle & $ O - C$\\
+2400000 & days  & & days 
\\[2mm]
\hline\\[-1mm]
48854.4006 & 0.0010 & 16140 & --0.0004 \\
48856.3625 & 0.0006 & 16167 & --0.0006 \\
49842.5264 & 0.0010 & 29740 & +0.0001 \\
50047.4171 & 0.0003 & 32650 & +0.0002 \\[2mm]
\hline
\end{tabular}
}
\end{quote}
\bigskip

The phase difference between our updated ephemeris and  that of Rea91 is
only 0.003 phase units at the time of the ROSAT X-ray observations presented
by Schlegel and Mukai (1995). Hence, if their X-ray light curve is shifted 
appropriately, it still displays the highest count rate around eclipse 
phase and  a dip 0.15 phase units in advance (at phase 0.85). 
While the dip may be explained
by the transient accretion stream, which is lifted out of the orbital plane,
there is no explanation for the missing X-ray eclipse. According to 
the standard model for X-ray emission in polars, there must be a 
X-ray eclipse present when an optical eclipse is observed. 
%The only
%exception requires a highly specific geometry and  is therefore unlikely.
%If \ep\ would be a two-pole accretor and  if one pole would preferentially
%emitting optical cyclotron radiation and  the other preferentiall
It seems that 
problems with the X-ray data analysis might have 
corrupted the light curve.

The peaks in the optical light curves are caused by strong cyclotron beaming,
hence, they contain information about the accretion geometry. With their
estimate of the orbital inclination $i = 80^\circ $, and  the observed length
of the bright phase, Rea91 have also derived
the probable location of the accretion spot, $\delta \sim 18^\circ $. 
Using this number at face value also for the inclination of the
field line in the spot, the angle between the field in the spot and  the
line of sight varies between $\theta_{\rm min} = i-\delta = 62^\circ $ and  
$\theta_{\rm max} = 90^\circ $ only. This range, in particular the 
rate of change of $\theta$ with phase, appears very small and  is likely
to be unable to produce the observed sharp intensity peaks. We have,
therefore, undertaken some simple light-curve modelling using 
a tabulated cyclotron model from Wickramasinghe and Meggitt (1985) for
$kT = 10$\,keV, size parameter $\log\Lambda = 7$ and  dimensionless
frequency $\omega/\omega_c = 12$ (see next section for a motivation of 
these numbers). The visibility and  radiation characteristic of a point-like
accretion spot are calculated for a given aspect of the observer, orientation 
of the magnetic field and  orbital phase. 

\begin{figure}[t]
\begin{minipage}[]{175mm}
\begin{center}
\begin{minipage}[]{80mm}
\psfig{figure=licus_ps,width=80mm,bbllx=27mm,bblly=36mm,bburx=145mm,bbury=249mm,
clip=}
\end{minipage}
\begin{minipage}[b]{14mm}
\end{minipage}
\begin{minipage}[]{80mm}
\psfig{figure=i80dvar_ps,width=80mm,bbllx=142pt,bblly=280pt,bburx=560pt,bbury=500pt,clip=}
\vspace{2cm}
\psfig{figure=lc_mod_dat_ps,width=80mm,bbllx=142pt,bblly=280pt,bburx=560pt,bbury=500pt,clip=}
\end{minipage}
\end{center}
\end{minipage}
\aba
Fig.~1a) (left).
Optical V-band light curves of \ep\ obtained at different occasions 
during recent years ({\it top:} AIP 70cm, {\it middle:} trailed spectrogram,
{\it bottom:} data depicted from Rea91). 
All light curves were normalized to an approximate
brightness level of 1 at phase \pe = 0.5. The data of 1995 and  1988 were 
phase folded, the light curve derived from the trailed spectrogram is shown 
in original time sequence.\\
b) (top right) Synthesized light curves of a cyclotron emitting 
accretion spot with zero extent. The orbital inclination was fixed at
$i=80^\circ $, the accretion spot colatitude was varied in steps of $20^\circ $
between $20^\circ $ (longest visibility) and  $80^\circ $
(shortest visibility). 
The phase convention defines minimum polar angle $\Theta$ as phase zero.\\
c) (bottom right)
Synthetic light curve assuming two cyclotron spots compared with the
1995 observations. 

\abe
\end{figure}

Light curves resulting from this 
model are shown for fixed inclination $i=80^\circ $ and  different colatitude 
in Fig.~\ref{ep_lcs}b. Even this simple model yields constraints 
for the accretion geometry from a comparison of the observed and  modelled 
width of the cyclotron peaks. The model light curves were calculated neglecting
any extent of the accretion spot in horizontal or vertical direction,
they are thus as sharp as theoretically possible. Any extent in either 
direction broadens the peaks.
The observed peaks are certainly affected by some extent of the accretion 
region. We, therefore, use for the comparison the sharpest of the observed 
peaks, the first one in 1992, which then gives a lower limit to the 
colatitude~$\delta$. The peak width is measured as full width at 20\%
of maximum intensity measured with respect to the faint phase intensity. 
The width from the 1992 light curve is 0.127, the numbers
for the synthesized light curves are given in Tab.~4.
There is some systematic inconsistency introduced since the background light
is time variable in the observed light curve, reaching a minimum around phase
0, while the background in the modelled light curve is constant. If we account
for a variable background the true width of the observed peak might be as 
large as 0.155 phase units.
%\ref{peak_par}. 
%\vspace{150mm}
%\vfill

%\newpage

%\bigskip

\begin{quote}
\aba
Table 4.
Full width of intensity peaks at 20\% level 
in synthesized light curves of Fig.~\ref{ep_lcs}b

\abe

\medskip

{\footnotesize
\begin{tabular}{rlrl}
\hline\\[-1mm]
colatitude & $\Delta\phi$ & colatitude & $\Delta\phi$\\[2mm]
\hline\\[-1mm]
20$^\circ$  &   --  & 60$^\circ$  & 0.109 \\
30$^\circ$  & 0.210 & 70$^\circ$  & 0.105 \\
40$^\circ$  & 0.157 & 80$^\circ$  & 0.099 \\
50$^\circ$  & 0.129 \\[2mm]
\hline
\end{tabular}
}
\end{quote}
\bigskip

The numbers given there show that the colatitude must be larger than 
40$^\circ - 50^\circ$.
With $\delta = 50^\circ $ the visibility of the spot is only 0.55 compared to 
0.64 as observed. This is clear evidence for an extended accretion region. Some
constraints on the likely shape of the extended region comes from the fact,
that the second peak in each of the observed 
light curves is broader than the first 
peak. This requires an elongated arc or ribbon which is better aligned with the
terminator of the line of sight at start of the bright phase than at end.
As a first approximation to the real situation we have modelled it
using two spots (points) which may be regarded as endpoints of the arc.
We again used $i= 80^\circ $, colatitudes $\delta_1 = 60^\circ $, $\delta_2 = 
70^\circ $, a phase difference between both spots of 10$^\circ$ \ and  a height
of both regions of 0.005\,R$_{\rm wd}$ (white dwarf radii). The resulting 
light curve is shown in comparison with our observed light curve of 1995
in Fig.~\ref{ep_lcs}c. Although not convincing in details (which is not 
surprising with the simple model assumptions), it reflects the main features 
of the observed light curve. It has the  correct length of the bright phase,
and the second peak is broader than the first.

At all occasions the light curves of \ep\ have the same length of the 
bright phase within 0.03 phase units. 
In addition, there is no significant shift of the bright 
phase with respect to eclipse center. Nevertheless, the length of the 
intensity peaks at start and  end of the bright phase and  their  relative 
brightness may change drastically. Several implications can be drawn from this 
observation. Firstly, the spin-orbit coupling of the white dwarf is probably 
higher than $1\times 10^6$ (inverse of fractional difference between the
spin and  the orbital periods). Secondly, the accretion arc has so far 
undergone no major migration towards different longitudes or latitudes.
This is perhaps somewhat surprising because of its extreme 
position in stellar longitude, $\psi \simeq -17^\circ $. The center of the
bright phase occurs after eclipse center, hence the magnetic axis points
at the trailing side of the secondary star and  thus resists the accretion 
torque which tries to turn the white dwarf to the leading side. Thirdly,
although the accretion arc, which we regard as a region where accretion 
potentially may occur, seems to be a fixed structure, the actual distribution
of matter (hot emission plasma) is highly variable, thus leading to different
height and  width of the cyclotron peaks. 

\begin{figure*}[t]
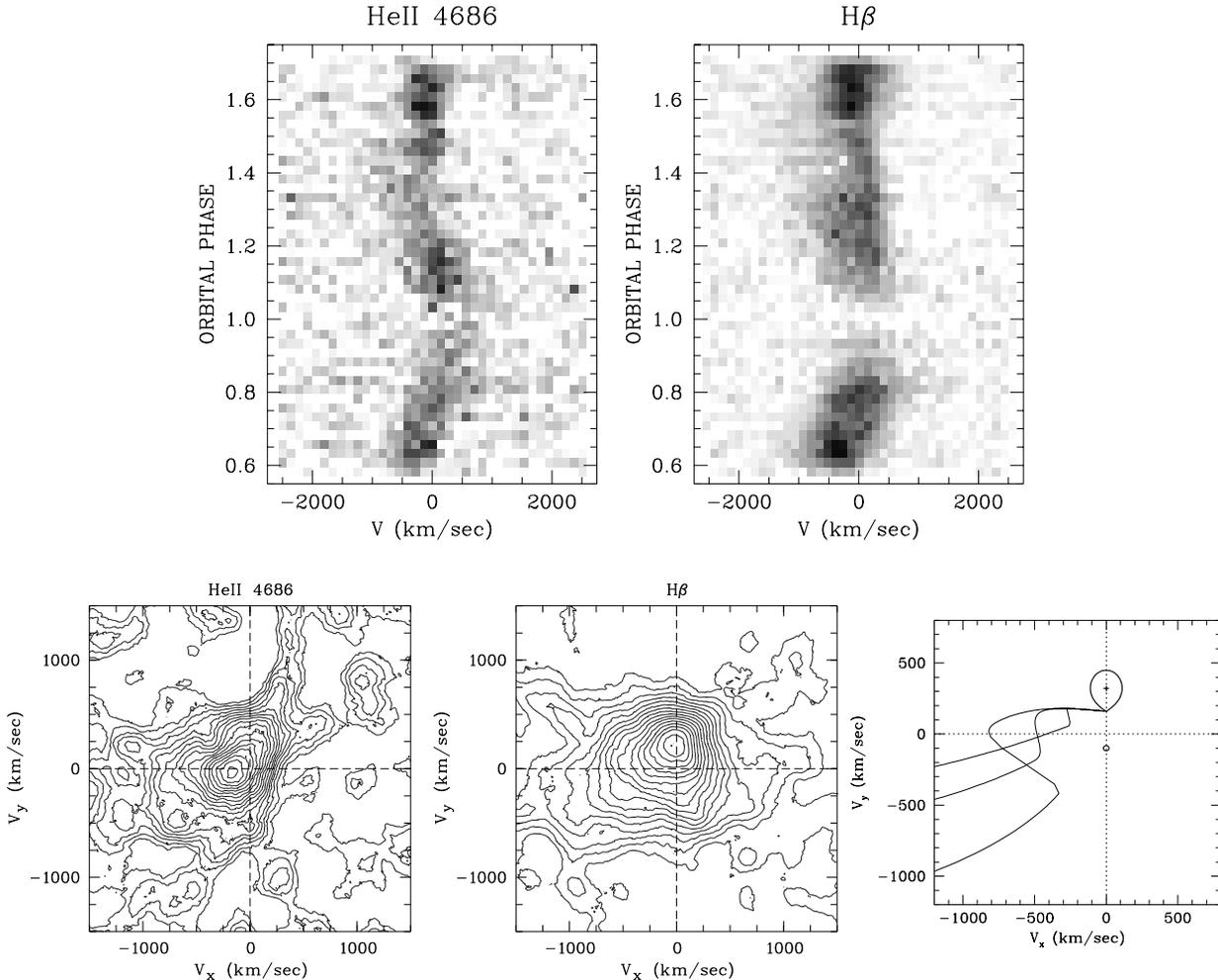

%\begin{center}
\begin{minipage}[]{188mm}
\begin{center}
\begin{minipage}[]{85mm}
\end{minipage}
\begin{minipage}[b]{25mm}
\end{minipage}
\begin{minipage}[]{85mm}
\end{minipage}
\end{center}
\end{minipage}
 
\vspace{0.5cm}
\begin{minipage}[]{188mm}
\begin{center}
\begin{minipage}[]{55mm}
%%%%%%%%%%%%%%%%5\psfig{figure=t3_he2_cont_ps,width=55mm,bbllx=26mm,bblly=28mm,bburx=183mm,bbury=188mm,angle=-90,clip=}
\end{minipage}
\begin{minipage}[b]{15mm}
\end{minipage}
\begin{minipage}[]{55mm}
%%%%%%%%%%%%%%%%5\psfig{figure=t3_hb_cont_ps,width=55mm,bbllx=26mm,bblly=28mm,bburx=183mm,bbury=188mm,angle=-90,clip=}
\end{minipage}
\begin{minipage}[]{55mm}
\end{minipage}
\end{center}
\end{minipage}
\caption[]{\label{toms}
Grey-scale representation of trailed, continuum-subtracted
spectra of the HeII $\lambda$4686 and  H$\beta$ lines of \ep .
Phase runs along the ordinate from bottom to top, wavelength has been 
transformed to velocity using the rest wavelengths of the specified lines.
In the lower panels the Doppler tomograms of the trailed spectra are 
shown computed by filtered backprojection.
{\it (bottom right)}
Locations of the secondary star and  the different parts of the accretion 
stream in the velocity-plane $(v_x, v_y)$ for an assumed mass ratio of 
$Q = M_1/M_2 = 3.2$. Shown are the centers of mass of both stars (on the 
axis $v_x = 0$), the shape of the secondary star, a ballistic trajectory 
starting at the $L_1$-point and  that part of the stream which is 
guided by the magnetic field. We assumed three arbitrary loci of the 
coupling radius, chosen typically for AM Her stars in general and 
likely to occur in \ep\ also.
}
\end{figure*}

A final comment in this section concerns the unpolarized
radiation observed while the accreting pole is behind the white 
dwarf. It was regarded as of ambiguous nature be Rea91, being either thermal 
emission from an extended column above the white dwarf surface or 
reprocessed emission from the heated face of the secondary star. The key 
to answer this question lies in the Doppler tomograms shown below 
(Fig.~\ref{toms}) and  the synthesized light curve of Fig.~\ref{ep_lcs}c. The 
former show line emission from the heated face of the companion star and 
the accretion stream and  the latter show a significant decrease of the 
bright-phase intensity with respect to the model curve during the nominal
bright phase outside the genuine eclipse. The intensity is there lower than 
during the nominal faint phase. Putting both things together, the likely 
nature of the unpolarized 
component is recombination radiation (line$+$continuum) from the accretion 
stream and  the heated face of the secondary. Both are well observable outside
the bright phase and  start their eclipse much earlier than the 
accretion spot on the white dwarf. A rough estimate of the 
corresponding phases can be derived using the opening angle of the cone
of the companion stars Roche lobe at the inner Lagrangian point $L_1$, which is
about 55$^\circ$. The self-eclipse of the heated face of the secondary starts
$90^\circ +55^\circ $ prior to eclipse center, hence at \pe$\simeq 0.6$, the 
eclipse of the accretion stream starts with eclipse of $L_1$, hence at 
\pe $\simeq 0.85$. These estimates show that large parts of the light curve
may be affected by shadowing and  obscuration of the very extended emission 
regions on the companion star and  the accretion stream.
 
\kap{Signatures of the magnetic field}
\begin{figure}[t]
\begin{center}
\begin{minipage}[]{120mm}
\psfig{figure=mean_spec_ps,width=12cm,bbllx=28pt,bblly=28pt,bburx=570pt,bbury=810pt,angle=-90,clip=}
\end{minipage}
\end{center}
\aba

Fig.~2. (upper panel) Mean bright- and  faint-phase spectra of \ep\ obtained 
in August 1992.
(lower panel) Difference of the above spetra, regarded
as cyclotron spectrum.

\abe

\end{figure}

\begin{figure}
\begin{center}
\begin{minipage}[]{120mm}
\psfig{figure=haloline_ps,width=12cm,bbllx=22mm,bblly=25mm,bburx=170mm,bbury=130mm,clip=}
\end{minipage}
\end{center}
\aba

Fig.~3. Mean trailed bright-phase spectrum of \ep\ centered on 
phase 0.8 showing H$\alpha$ Zeeman absorption troughs. The model below 
the observed spectrum was calculated for a Gaussian distribution of the
magnetic field with central field strength of 16\,MG and  spread $\sigma_B$ of
1\,MG.

\abe

\end{figure}

\begin{figure}[t]
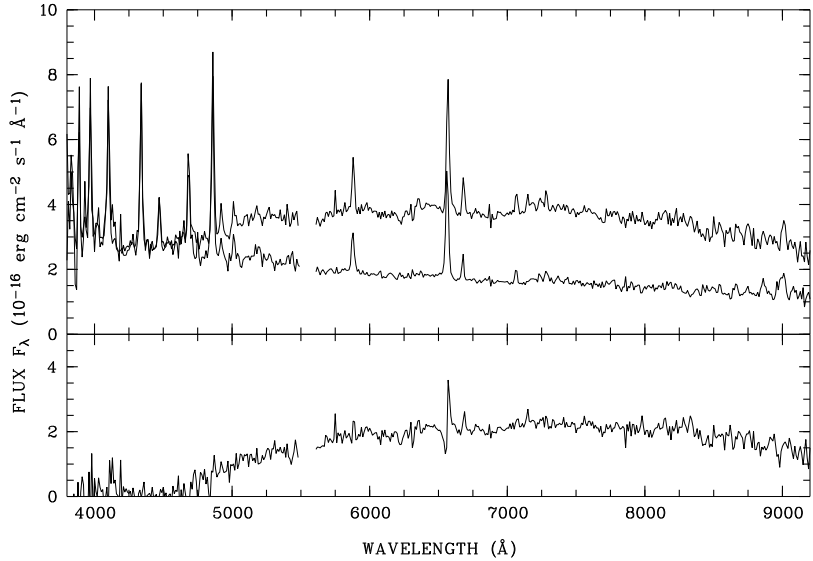

\caption{\label{meanspec} 
{\it upper panel:} Mean bright- and  faint-phase spectra of \ep\ obtained 
in August 1992. {\it lower panel:} Difference of the above spetra, regarded
as cyclotron spectrum.
}
\end{figure}

In Fig.~\ref{meanspec} we show mean bright- and  faint-phase spectra of 
\ep\ obtained in August 1992. They show the typical features of an AM Her star
with strong H-Balmer and  HeI, HeII emission lines superposed on a 
non-photospheric continuum. The continuum rises to the blue 
during the faint phase,
it has maximum light in the red during the bright phase. 
On the assumption, that the excess light
in the bright phase  is only due to cyclotron radiation, a cyclotron spectrum
can be computed by subtracting the mean faint-phase spectrum from the
mean bright-phase spectrum. The difference spectrum 
is shown in the lower panel of 
Fig.~\ref{meanspec}, it is very red with peak intensity at $\sim$7000\AA\ 
and with no important contribution shortward of $\sim$5000\AA. 
%\newpage
%\vspace*{82mm}

%\vspace{93mm}

%\bigskip

\begin{figure}
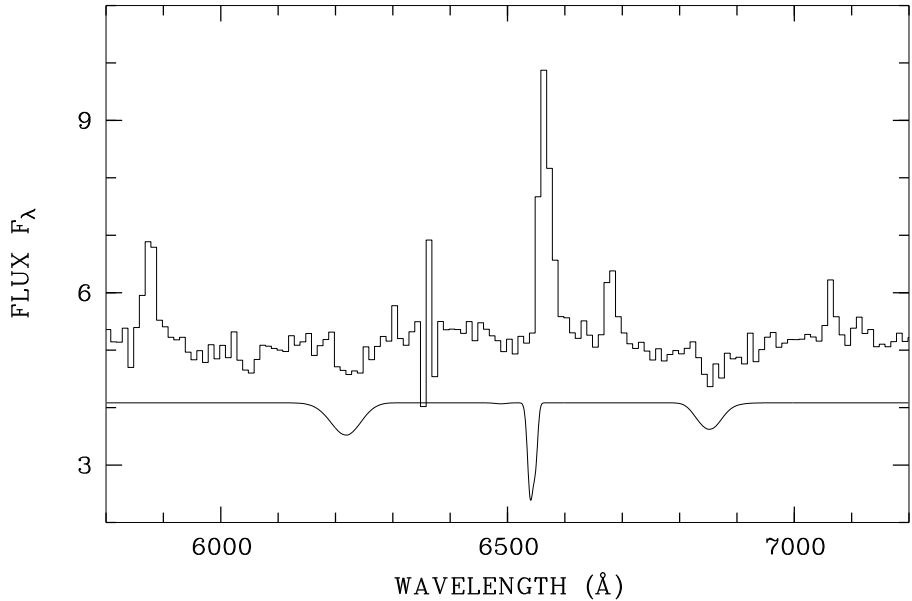

\caption{\label{haloline} 
Mean trailed bright-phase spectrum of \ep\ centered on 
phase 0.8 showing H$\alpha$ Zeeman absorption troughs. The model below 
the observed spectrum was calculated for a Gaussian distribution of the
magnetic field with central field strength of 16\,MG and  spread $\sigma_B$ of
1\,MG.
}
\end{figure}

The bright-phase spectrum shown in Fig.~\ref{meanspec} was computed by 
averaging the stepped spectra before and  after the eclipse centered on
phases 0.790 and  0.277, respectively. A closer look on the trailed spectrum
at around phase 0.8 is shown in Fig.~\ref{haloline}. It displays two distinct
flux minima more or less symmetric to the H$\alpha$ emission line. Such 
features have 
been found in several polars and  were interpreted as Zeeman-split
H$\alpha$ absorption lines in a cool halo surrounding the or intermixed
with the hot emission plasma. This explains why the Zeeman lines are
visible only against the bright cyclotron background and  not during
phases of low cyclotron brightness. We take the same view here and  estimate
the field strength in the halo to $B = 16$\,MG. Below the observed spectrum
we show the run of averaged Zeeman-split absorption components calculated 
assuming a Gaussian spread with $\sigma_B = 1$\,MG around the central field
of 16\,MG. The absorption troughs of the $\sigma^\pm$-components 
are reflected well by this simple model, the $\pi$-component is filled 
in significantly by the strong emission line.

%\newpage

%\vspace*{145mm}

%\bigskip

Three polars are known which show both, cool halo Zeeman absorption lines and  
hot plasma cyclotron emission lines (MR Ser, V834 Cen, and  DP Leo, see
Schwope (1995) for a complete compilation of measured field strengths in 
polars). In all three cases very similar values  of the field strengths for
both features were measured indicating the close coexistence of cool 
infalling matter and  hot settling plasma and  qualifying halo lines as 
tracers of the field strength in the accretion region. There is one possible
counterexample, BL Hyi, showing 12\,MG halo lines (Schwope et al.~1995) 
and somewhat noisy 18\,MG cyclotron lines (Ferrario et al.~1996). 
If we assume a field of 16\,MG to be present in the cyclotron region too, the
peak intensity of the observed cyclotron spectrum can be reflected with 
an isothermal 10 keV 
cyclotron model (see e.g.~Schwope et al.~1995) with optical
depth parameter $\log \Lambda \simeq 7$. At a field of 16\,MG the spectral 
range between 5000\AA\ and  9000\AA\ corresponds to harmonic numbers $8.3-14.9$
for a 10 keV plasma. In this  regime significant harmonic overlap exists 
and we do not expect individual harmonics resolved, as observed. 
One might be tempted to identify some intensity humps in the bright-phase
spectrum of Fig.~\ref{meanspec} (e.g.~at $\lambda\lambda$5900, 6500, 7300\AA) 
with individual cyclotron harmonics, but this seems
to us a combined effect of residual low-frequency fringing in the 2D-CCD 
spectrum, halo Zeeman absorption and  a somewhat uncertain response function 
in the vicinity of the beam-splitting wavelength.

\kap{Emission lines and  stellar masses}
\begin{figure}[t]
\begin{minipage}[]{175mm}
\begin{center}
\begin{minipage}[]{85mm}
\psfig{figure=he2_tr_ps,width=55mm,bbllx=17mm,bblly=65mm,bburx=132mm,bbury=218mm,clip=}
\end{minipage}
\begin{minipage}[b]{25mm}
\end{minipage}
\begin{minipage}[]{85mm}
\psfig{figure=hbet_tr_ps,width=55mm,bbllx=17mm,bblly=65mm,bburx=132mm,bbury=218mm,clip=}
\end{minipage}
\end{center}
\end{minipage}
 
\vspace{0.5cm}
\begin{minipage}[]{175mm}
\begin{center}
\begin{minipage}[]{55mm}
\psfig{figure=t3_he2_cont_ps,width=55mm,bbllx=26mm,bblly=28mm,bburx=183mm,bbury=188mm,angle=-90,clip=}
\end{minipage}
\begin{minipage}[b]{15mm}
\end{minipage}
\begin{minipage}[]{55mm}
\psfig{figure=t3_hb_cont_ps,width=55mm,bbllx=26mm,bblly=28mm,bburx=183mm,bbury=188mm,angle=-90,clip=}
\end{minipage}
\begin{minipage}[]{55mm}
\psfig{figure=sketch_tom_ps,width=50mm,bbllx=28pt,bblly=190pt,bburx=370pt,bbury=500pt,clip=}
\end{minipage}
\end{center}
\end{minipage}
\aba

Fig.~4.
Grey-scale representation of trailed, continuum-subtracted
spectra of the HeII $\lambda$4686 and  H$\beta$ lines of \ep .
Phase runs along the ordinate from bottom to top, wavelength has been 
transformed to velocity using the rest wavelengths of the specified lines.
In the lower panels the Doppler tomograms of the trailed spectra are 
shown computed by filtered backprojection.\\
(bottom right)
Locations of the secondary star and  the different parts of the accretion 
stream in the velocity-plane $(v_x, v_y)$ for an assumed mass ratio of 
$Q = M_1/M_2 = 3.2$. Shown are the centers of mass of both stars (on the 
axis $v_x = 0$), the shape of the secondary star, a ballistic trajectory 
starting at the $L_1$-point and  that part of the stream which is 
guided by the magnetic field. We assumed three arbitrary loci of the 
coupling radius, chosen typically for AM Her stars in general and 
likely to occur in \ep\ also.

\abe
\end{figure}

Our trailed spectrograms of \ep\ contain important information about the
emission lines although the spectral resolution is rather low, 6\,\AA\ FWHM. 
The low spectral resolution is somewhat compensated by the good 
phase resolution. As two examples we show in Fig.~\ref{toms} the trailed 
spectra of the HeII $\lambda$4686 and  H$\beta$ lines together with 
their Doppler images. The latter were computed by a filtered backprojection
of the spectral lines. The spectra were continuum-subtracted for backprojection
and representation in Fig.~\ref{toms}. 
Both lines have in common that they are bright before and  after eclipse and  
that they show a flux depression between phases 0.35 and  0.5. The eclipse
seems to be total for one or two phase bins, although definite statements
cannot be drawn due to the low signal-to-noise ratio there. From phase 0.6
to 0.9 both lines show a pronounced motion from blue- to redshifts, between
phases 0.05 and  0.35 in the opposite direction. This is naturally explained
by emission from the accretion stream, which points away from us during 
eclipse and  shortly thereafter. 
This stream component is rather broad, the measured width of $\sim$
15\,\AA\ FWHM corresponds to a velocity dispersion of $\sim$900\,km s$^{-1}$.
The Balmer lines show in addition an unresolved line of much lower radial
velocity amplitude which is most prominent
at superior conjunction of the secondary star (\pe $\simeq 0.5$). It cannot 
be traced in the trailed spectrogram of the HeII line. 
This unresolved line is reminiscent of 
the narrow emission line of reprocessed origin from the secondary star, being
so prominent in HU Aqr (Schwope et al.~1996). The Doppler image of H$\beta$
supports this impression. If we compare the Doppler images of He{\sc 
ii}~$\lambda$4686 and  H$\beta$, it becomes clear, that the line 
emission in H$\beta$ is much more concentrated on the secondary star
(or the region of the inner Lagrangian point $L_1$) and  that the He{\sc 
ii} emission comes predominantly from the accretion stream. A sketch of 
possible locations of emission line regions, facilitating an understanding
of the Doppler maps, is given also in Fig.~\ref{toms}.

Interestingly, the Doppler image of HeII\,$\lambda$4686  is clearly 
different from that seen in HU Aqr, suggesting a much denser accretion 
curtain in \ep\ than in HU Aqr, which provides sufficient shielding 
of the companion star by absorbing ionizing EUV-photons. 

We have measured the positions of the narrow emission lines of H$\beta$ and 
H$\gamma$, whose trailed spectrograms look very similar, in the stepped
spectra. This was done by fitting a double-Gaussian (narrow plus broad line) 
with fixed width of the narrow line. An unweighted sine fit to the measured 
positions in the phase interval 0.25--0.75, where both lines could be 
separated from each other, yielded a radial 
velocity amplitude of $210\pm25$\,km/sec. This represents the center of line
emission from some location on the illuminated hemisphere and  can be used
for a mass estimate of the white dwarf if some assumptions are fulfilled:
1) emission happens over that part of the Roche lobe which is geometrically 
accessable by radiation from the white dwarf, 2) ionizing radiation is 
reprocessed and  re-emitted locally, 3) the secondary is a main-sequence 
star. We estimate the mass of the secondary star using the late-type star
mass-radius relation given by Neece (1984) to be $M_2 = 0.133$\,M$_\odot$.
With (1) and  (2) a geometrical reprocessing model (Beuermann and Thomas 1990) is
applicable in order to transform center-of-light to center-of-mass radial
velocities. This results in a mass ratio $Q = M_1/M_2 = 3.2 \pm 0.5$ and 
in a white dwarf mass of $M_1 = 0.43 \pm 0.07$\,M$_\odot$. These results are
only slightly dependent on the true value of the inclination, we used $i = 
80^\circ $, but on the details of the illumination and  the true mass-radius 
relation (which is rather uncertain). If the emission is concentrated
towards the inner Lagrangian point $L_1$, our value of $M_1$ would represent 
a lower limit only. However, taken the face values, the radial velocity 
amplitude gives evidence for a somewhat undermassive white dwarf. 

\kap{Concluding remarks}
We have presented the first phase-resolved spectroscopic and  new 
photometric data of the eclipsing polar \ep. Using these we could improve on
the accuracy of the eclipse ephemeris. We found no evidence for a quadratic 
term to be present in the ephemeris 
nor significant shifts of the bright phase
with respect to eclipse center over a seven-year basis.
Our models for the optical light curves reveal evidence for an extended 
accretion arc located at a stellar colatitude of about $60^\circ$--$70^\circ $.
The X-ray light curve presented by Schlegel and  Mukai (1995) remains
ambigious.

We have identified H$\alpha$ Zeeman absorption lines originating in an
accretion halo at a field strength of 16\,MG, which is a relatively low value
compared to other AM Her stars. The red colour of the cyclotron spectrum
is also suggestive of a relatively low field strength.

The trailed spectrograms of high- and  low-ionization lines 
(e.g.~HeII\,$\lambda$4686, H$\beta$) 
are clearly different as well as the Doppler maps constructed
from them. Balmer line emission is more concentrated on the heated face
of the secondary and  the region around $L_1$, whereas stream emission 
is more prominent in HeII\,$\lambda$4686. 
A possible explanation for the weakness of 
HeII\,$\lambda$4686 
emission from the secondary star is the presence of an accretion 
curtain providing effective shielding. If so, also the H-Balmer line is
likely affected. Our radial velocity measurement of the secondary would
probably be too low as well as our present estimate of the white-dwarfs mass,
$M_1 = 0.43\pm0.07$\,M$_\odot$.

{\it Acknowledgements.}
We thank P.~Notni for invaluable assistance 
at the AIP 70cm telescope and  J.-U.~Fischer,
I.~Lehmann and  A.~Sch\"{u}ller for patient observing during cold 
November nights in Potsdam. We acknowledge useful comments of the referee
K.~Beuermann.
This work was supported by the DFG under grant Schw536/1-1 and  
the BMB+F under grant 50 OR 9403 5.

%
%%%%% References
\refer
\aba

\rf{ Beuermann K., Thomas H.-C.: 1990, 1990, Astron. Astrophys.~230, 326
}\rf{ Ferrario L., Bailey J., Wickramasinghe D.: 1996, Mon. Not.
R. Astron. Soc.~282, 218 
}\rf{ Neece G.D.: 1984, Astrophys. J.~277, 738
}\rf{ Remillard R.A., Stroozas B.A., Tapia S., Silber A.: 1991,
Astrophys. J.~379, 715 (Rea91)
}\rf{ Schlegel E.M., Mukai K.: 1995, Mon. Not. R. Astron. Soc.~274, 555
}\rf{ Schwope A.D.: 1995, Rev.~Mod.~Astron. 8, 125
}\rf{ Schwope A.D., Beuermann K.: 1989, Astron. Astrophys.~222, 132
}\rf{ Schwope A.D., Beuermann K., Jordan S.: 1995, Astron. Astrophys.~301, 447
}\rf{ Schwope A.D., Mantel K.-H., Horne K.: 1996, Astron. Astrophys., in press
}\rf{ Wickramasinghe D.T., Meggitt S.M.A.: 1985, Mon. Not. R.
Astron. Soc.~214, 605} 
\rf{}

\abe

%%%%% End of references
%
%%%%% Address of the authors
\address
Axel D. Schwope, Sabine Mengel\\
Astrophysikalisches Institut Potsdam\\
An der Sternwarte 16\\
D--14482 Potsdam\\
Germany\\
e-mail: ASchwope@aip.deEND
%
%%%%% End of address

\newpage

\end{document}